# A Protocol to Convert Infrastructure Data from Computer-Aided Design (CAD) to Geographic Information Systems (GIS)

Eric Sergio Boria*[1], Mohamed Badhrudeen*, Guillemette Fonteix^, Sybil Derrible*, Michael Siciliano*

* University of Illinois at Chicago (UIC)
^ École Nationale des Sciences Géographiques (ENSG), France

## Abstract

While many municipalities and organizations see value in converting infrastructure data from Computer-Aided Design (CAD) to Geographic Information System (GIS) format, the process can be complex, expensive, and time-consuming. Given that municipal employees often prefer to continue performing work in both CAD and GIS, depending on the type of work required, an improved conversion process would help municipalities more fully employ GIS-based analyses. Municipalities facing budget and capacity challenges would especially benefit from an improved conversion process. With advances in GIS functionality and the promise of smart and connected cities, more emphasis is placed on the quality of data, and in this case, the potential loss of data quality from CAD to GIS formats. The goals of this article are twofold. First, to understand the common practices municipalities use to convert infrastructure CAD data to GIS and the specific challenges they face. Second, based on knowledge of those practices and challenges, this article proposes a five-step process to reduce common conversion errors and reduce the time required to correct these errors. The process is illustrated through the conversion of CAD data from the University of Illinois at Chicago (UIC) campus. The findings were validated with qualitative, semi-structured interviews conducted with GIS Analysts and Managers working in eleven municipalities across the United States who directly manage at least one of the following infrastructures: water, sanitary sewer, or stormwater sewer systems. The interviews confirmed the challenges municipalities faced with the conversion and identified solutions interviewees undertook to enable data-informed decision-making.

---

[1] Corresponding author at: University of Illinois at Chicago, CUPPA Hall, 412 S. Peoria, Chicago, IL 60607
Email address: eric.boria@protonmail.com (E. Boria)

## Acknowledgements

The authors would like to thank the National Science Foundation for supporting this work and the interviewees from the eleven participating municipalities for giving us the time and sharing their experiences for this study. Also, we would like to thank the University of Illinois at Chicago (UIC) Office of Capital Planning & Project Management (OCPPM) for taking their staff time to participate and for providing the CAD data used in this study.

# 1. Introduction

The effective management of utility services, such as water and wastewater, rely on an accurate assessment of the systems themselves. For such an assessment, correct and complete maps are crucial. However, the quality and format of these maps vary considerably. Many of the municipal employees interviewed reported that they continue to utilize Computer-Aided Design (CAD) and paper maps. While data in Geographic Information Systems (GIS) format can provide additional functionality and has become the dominant digital format used for mapping, some employees prefer to conduct design work in CAD. Many of the layers of information that can be combined in GIS cannot be adequately incorporated in a CAD system (Bansal 2014). GIS possesses several benefits that make it a superior format as compared to CAD. First of all, GIS allows for geospatial analysis, while CAD is purely a digital drawing format. Moreover, mapping data with GIS is one way to create geospatial data, which has been used to effectively maintain and upgrade critical infrastructures (Kulawiak and Lubniewski 2014). Furthermore, GIS can be easily combined with different modeling environments to perform scenario analysis in the context of infrastructure planning and management (Pior and Shimizu 2001; Samela et al. 2018), for example in finding areas that are vulnerable to flooding (Kermanshah, Derrible, and Berkelhammer 2017; Wisetjindawat et al. 2017).

This study includes interviews with municipal employees (see Section 2), most of whom stated that one of the reasons for converting the infrastructure data into GIS was to have centralized access to the combined infrastructure data. Centralized access allows employees in various roles from public works crews to planners to better maintain, repair, and upgrade infrastructures with a lower risk of accidental failures. The quality of geospatial data is valuable due to its use in visualization for excavation (Talmaki, Kamat, and Cai 2013). Data from existing buildings that have been included on as-built drawings (changes that are marked as a building or line is created) can be messy or missing, which can lead to difficulties in managing the data (Lu and Lee 2017). Moreover, the Open Geospatial Consortium (OGC) report concluded that centralized utility data organized according to standardized, geospatially enabled data models could enable interoperability between infrastructure systems (Lieberman and Ryan 2017). More generally, the presence and availability of GIS data for multiple infrastructure systems (e.g., electricity, gas, water, wastewater, transport, and telecommunications) also allows for a study of infrastructure interdependencies, which can contribute to the general body of work for smart, sustainable, and resilient cities (Derrible 2017; 2018; Mohareb, Derrible, and Peiravian 2016).

Generally, the potential benefits of managing underground infrastructure data in GIS motivate public and private organizations to undertake the conversion of various data types to a GIS format. Among the municipalities that participated in the study, the conversion of existing infrastructure information into a GIS format was often conducted as each municipality saw fit, which indicates the lack of formal standards for the conversion process (Balasubramani et al. 2017). Other studies that included interviews with building professionals have also reported that the lack of standards hinders the widespread use of new technologies (Castronovo et al. 2014). Many of these municipalities experience budgetary and capacity challenges. Hence, it is beneficial to have a systematic process that would allow municipalities to more quickly convert the data format from CAD to GIS while maintaining data quality. At the time of this writing, the mainstream mapping and analysis software packages used for GIS are ArcGIS® by the ESRI (Environmental Systems Research Institute) Corporation and the free and open source GIS software package Quantum GIS (QGIS). AutoCAD® and Microstation software packages were the most used for engineering drawings. The municipalities interviewed for this study commonly used these software packages.

CAD is a digital format that renders technical illustrations of objects in the real world, and it is also applied to problems of design, including in the civil, mechanical, and electrical engineering disciplines. For this reason, there are multiple attempts to convert CAD data to other formats and systems that have a broader functionality, such as a building information model (BIM) (Yang et al. 2020). Unlike CAD, GIS data is stored in spatial databases, which allow users to manipulate information in relation to other information seamlessly. Another aspect of GIS data is the inclusion of topology, which describes the relationship between adjacent features (Law and Collins 2015). GIS brings these objects together into a logical group. By contrast, CAD is typically used for individual designs and is rarely used as a stand-alone database to manage multiple drawings. In CAD, objects have no relationships between one another. Another disadvantage of using CAD is that it does not allow the user to perform spatial analysis. This would be particularly important for municipalities interested in combining spatial datasets, such as comparing areas with aging infrastructures prone to leakage with data on the income level of residents. GIS has this ability to do spatial, topography, analysis (Law and Collins 2015). Besides, GIS follows different rules than CAD for drawing objects and archiving data of urban infrastructure systems. With the conversion of data from CAD to GIS, spatial information and attributes are an enhancement upon CAD drawings; this further enables "georeferencing" the entities that can provide more relevant details and data precision.

Despite the many benefits of converting CAD to GIS data, the conversion itself can be significantly challenging and often leads to many data inaccuracies. In the literature, few studies have looked into technical issues related to the conversion of data formats from CAD to GIS or provided approaches on how to deal with them. A study by Xie et al. (2015) delineated the stages that would minimize the loss of information during the conversion process. The stages include pre-analysis, conversion, and adjusting (Xie et al. 2015). Another approach to the conversion process explored interoperability between CAD and GIS (Peachavanish et al. 2006). He et al. (2011) found that coordinate transformations and features distortions are some of the common problems that need to be addressed during the conversion process. Further, several third-party software packages are available for the conversion and provide GIS shapefiles as output but are prone to other problems such as not being projected to the correct coordinate system, thus introducing feature distortions (Zhen, Jing, and Chen 2012). Finally, although protocols exist to convert GIS data into a network, they do not address CAD to GIS conversion issues (Karduni, Kermanshah, and Derrible 2016).

To address this gap in research, the main goals of this article are twofold. First, they are to gain insights into the practical problems associated with CAD to GIS conversion by conducting interviews with municipalities and reporting the main findings. Interviews with practitioners have been shown to reveal issues with the implementation of technology and novel recommendations from the field (Castronovo et al. 2014). Second, they are to propose a formal protocol to convert CAD data to GIS that can reduce the amount of time spent on the conversion process and the loss of information. The conversion process is explained using a case study to convert CAD data of the underground stormwater pipe network of the University of Illinois at Chicago (UIC) campus to GIS format. Previous research utilizing a campus case study approach has shown that GIS is used in spatial planning to assess the impact of any new facility upon its surroundings (Bansal 2014).

In the next section, we give a brief overview of the common challenges experienced by the municipalities interviewed for this study. Section 3 presents and details the proposed five-step process. Section 4 illustrates the process using the UIC case study. The article ends with a summary and suggestions for future research.

## 2. Challenges experienced by practitioners

Exploratory interviews were conducted to understand the challenges municipal GIS departments experience in converting CAD to GIS. These interviews provided insight into which steps in the process were the most time consuming. These time constraints make it challenging for municipalities to develop complete and accurate GIS-based maps, especially in departments with limited resources and staff capacity. With accurate and complete maps, municipalities and other organizations would be able to take full advantage of capabilities enabled by GIS-based data.

### 2.1. Methodology

Qualitative, semi-structured interviews were conducted with GIS analysts and managers working with eleven municipalities across the United States (U.S.). Given that the objective was to learn from the practitioners directly, the questionnaire was designed to explore what challenges they experienced and how they went about resolving those issues.

2.1.1 Site selection

Sites were selected from a list of all municipalities in the U.S. with a population of at least 10,000 residents. A random number generator was applied to the list to generate a sample of 40 municipalities. The justification for this population requirement is the assumption that municipalities need a minimum population, as a proxy for the municipal budget, to support the hiring of GIS-trained personnel. To be included, the municipality was also required to have control over managing at least one of the following infrastructures: water, sanitary sewer, or stormwater sewer.

2.1.2 Participant selection

Interviews were requested with GIS directors, managers, and analysts within the selected municipalities. To be included in the study, participants were required to have some experience with converting data from CAD to GIS and work with data from water, sanitary sewer, or stormwater sewer systems.

2.1.3. Description of participants

Out of the 40 municipalities contacted, interviews were conducted with 11, for a response rate of 28%. The distribution of municipalities by region and population is as follows:

| Region | Municipalities |
|---|---|
| Northeast | 1 |
| Midwest | 3 |
| South | 0 |
| Mountain | 3 |
| Southwest | 2 |
| West | 2 |

| Population (in thousands) | Municipalities |
|---|---|
| 10-20 | 3 |
| 20-30 | 2 |
| 30-40 | 1 |
| 40-50 | 1 |
| 50-60 | 1 |
| 60-70 | 0 |
| 70-80 | 1 |
| 80-90 | 1 |
| 90-100 | 0 |
| >100 | 1 |

Table 1: Distribution of participating municipalities

## 2.2. Motivations for converting data to GIS

To understand how municipal employees convert infrastructure data from CAD to GIS, it first requires asking why such a conversion is necessary. The following interview quote indicates a reason:

"So [the engineers] go out and GPS the underground stuff, which is what they really want to know where it is. They take that and they bring it into CAD cause that's what they're familiar with. And then they shipped the CAD files over to us [the GIS Department] and we bring it into our GIS and also into our asset management system." (Interview 005 July 12, 2018).

Interviewees stated that certain functions, such as building design, can be done more easily in CAD. A second reason, as stated in the above quote, is that those familiar with CAD prefer to work in that format, which then requires converting to GIS.

Municipalities often have an internal champion who encourages staff to transition data to a GIS format, hire GIS analysts and managers, and in some cases, have created GIS departments. Some of the reasons given for the emphasis on GIS-based data are stated in the following quotes:

"So it's for better accessibility for everybody else who's in the municipality to access that data." (Interview 004, July 11, 2018)

"The main benefit of converting [data] to GIS is you work with your asset management system. And retrieval is much easier for an online mapping." (Interview 005, July 12, 2018)

"I think you realized that many cities struggle, us included, in terms of how to handle all of that information. There is a component of making interactive maps that's becoming more prevalent and it's nice to have that for some of the field work, but we still extensively make static maps, whether that's in paper form or creating a PDF that's going to go out to people and being able to do that. It's difficult to in CAD, I know they can make some maps but they just don't have the same accessibility in terms of being able to read them and get information off of them." (Interview 011, July 19, 2018)

The interviews also revealed that some municipalities give more importance to the completeness of infrastructure data in maps than the accuracy of their locations. The reason being that engineers and other Public Works employees in the field primarily need to know what is supposed to be underground where they plan to dig. Then, they will use their standard practices to survey the location and find the precise locations of relevant structures. Interviewees spoke about the collective experience of those working in the field to know the details of underground structures gained through years of doing the work. In addition to the analysis and presentation of data in GIS, a common theme underlying the process of converting data to GIS is to create a systematic process to maintain, build upon, and improve the knowledge of the field, as summarized in the following quote:

"I think the short-term goal is to get all the information that is in a few of the older employees' heads and memories into the PC [personal computer] and GIS. So that when they leave, when they retire, that we're not going to lose all that information." (Interview 011, July 19, 2018)

**2.3. Reported challenges to the conversion process**

The challenges interviewees reported can be categorized into three main categories: incomplete data, inaccurate data, and conversion issues. These challenges are summarized in Table 2 below.

2.3.1 Incomplete data

The challenges municipal employees face in the conversion of infrastructure data from CAD to GIS was not always the technical process of conversion. Many of the stated problems derived from the incompleteness and inaccuracies in the data submitted. This may arise from a difference in understanding what data is necessary to collect for decision making. For example, as evidenced in the following quote, data collection should be aligned with decision making needs. However, if those collecting the data in the field are unaware of what other departments may need specific data for, that data may not be collected. Thus, the first step in the conversion process is to identify the data the municipality needs and set up a process to encourage the collection and submission of that data.

> "a lot of the prioritization comes from working with the engineering department in terms of what they have said that they want to look at in the future as well as working with our utilities department to see what attributes they're interested in as well for maintenance and … what information is useful for them when they're in the field." (Interview 011, July 19, 2018)

2.3.2 Inaccuracies in the original data

Even if the data existed in the original CAD files, there could have been errors in how it was initially recorded. The following description exemplifies how inaccurate data can end up in CAD and subsequently introduce errors into the GIS database:

> "the old stuff, … the old pipes in the ground for a long time. When we originally collected the data, we had an intern go around with one of the older guys that [has] been there a while and he'd say, well this is a six inch and I think it's AC and it was put in and the 60s, so it would get an install date of 1960. So, the older stuff was kind of 40, 50, 60, 70, until we really start to nail down when the actual installation day was. So that the problem was that they didn't quite remember correctly and they say, I think it's right here. So, we draw it in here and then they'd go to dig it up and starting digging sideways until they actually found it. So, the locations were off because the data wasn't kept. And then … the sizes were different than what they thought they were. the pipes were occasionally different than what they thought they were also." (Interview 005, July 12, 2018)

2.3.3 The conversion process

The steps that require the most time and present the most difficulties to those interviewed were often structuring attributes (georeferencing). Much of this data was either not converted from CAD or not collected even for the CAD drawing. The GIS managers reported that their preferred solution was to send workers into the field to check the accuracy of the data and insert those verifications or updates directly into GIS.

| Challenges with Data | Challenges with Conversion |
|---|---|
| Incomplete data | Attribute structuring |
| Inaccurate data | Topology |
| Data collected does not match need | Inconsistent naming practices |

Table 2: Summary of challenges interviewees reported

**2.4. Solutions employed and recommended**

Common themes emerged to interviewees' responses to what changes they believe would improve the accuracy, completeness, and practical use of GIS-based infrastructure data. The responses are summarized into the following three types of standards: naming convention, as-built drawings, and process.

2.4.1 Standard naming convention

To ensure that structures would be marked similarly from one municipality or organization to the next. This consistency would facilitate the sharing and use of locally-produced GIS maps, as stated by one interviewee:

> "If other cities would ask us for data, we would definitely share with them. They're working on a project that shares borders on our boundary. That happens often - sharing data with other developers or engineering firms. And sharing that data via GIS and shapefiles or geodatabases work best rather than through CAD." (Interview 011, July 19, 2018)

2.4.2. Standards for as-built drawings

To ensure the delivery of complete and accurate data to be converted into GIS. This would apply to contractors, developers, and engineers who provide the construction data to the GIS department. One interviewee expressed this request as such:

> "We would have to … push it back onto developers where they would be responsible for providing the Autocad [files] and then they would also provide shapefiles, feature classes of what we're interested in, and we would define what those are, and then they'd be providing all of that data input already and a geodatabase template that has domain feature classes all set up with all the data that we want them to fill in." (Interview 011, July 19, 2018)

2.4.3. Standard process

Ensuring that the data is treated in the same way will ensure consistency in GIS data quality across locations. An example of interviewees finding solutions to conversion challenges is seen in the following description where a manual sampling of data points was employed to resolve topology problems:

> "Yes, we have had topology problems, especially when we started working with geometric networks. There's all kinds of other issues that we've run into. They've tried a bunch of different things within CAD or engineering texts in terms of trying type network set of things and it's just inconsistent for getting everything all linked together. So, there's a certain amount of fixing that we will do. We'll import that data and usually project my project so it's not a huge extent of data. So, it will come in, we'll ensure that it's in the correct location. Sometimes they'll have forgotten to correctly project the data, so we'll have to send it back to them to get it projected in CAD and then we'll import it to our GIS and we'll look at it. And if it's not hooking into our existing line networks, we'll manually just attach it to the known networks, just to ensure that it's kind of taking care of some of that stuff. So, it's, inspected manually, but you know, it's usually two or three spots where you have to connect it into existing networks." (Interview 011, July 19, 2018)

The issue arises if each GIS analyst, or municipality, uses their own, different solutions to such problems. This will result in slight differences in GIS-based maps from one municipality or organization to the next. Thus, the need arises for a standard process. The following section proposes one approach to developing a standard process for CAD to GIS conversion.

## 3. Conversion Protocol

In this work, we propose the five-step process shown in Figure 1 and detailed in this section.

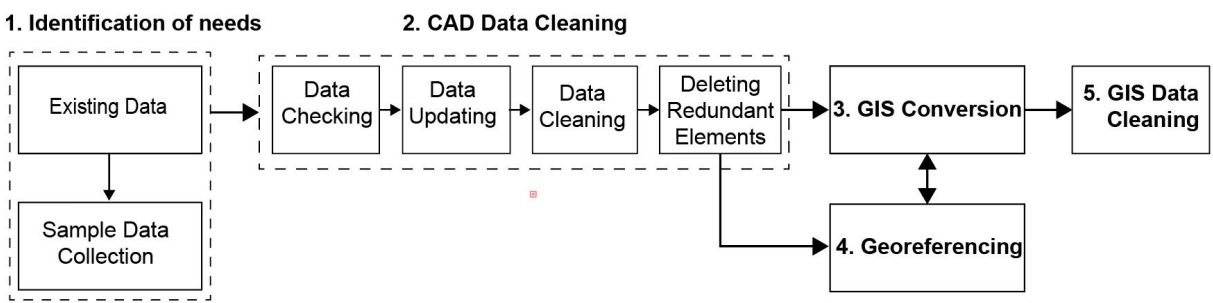

Figure 1: Proposed five-step process

### 3.1. Proposed five-step protocol

3.1.1. Identification of Needs

The information defined in Step 1 derives from the relevant actors, in this case, municipal departments, to identify the information to be collected. The *needs identification* step shapes the subsequent steps. It includes the collection of field data regarding natural and constructed infrastructure systems. The interviews with municipal GIS managers revealed a wide diversity of data types collected by municipalities. While some municipalities are advanced in establishing GIS departments and have procedures in place to upload data in GIS format on municipal infrastructures like water distribution, others collect CAD drawings, which was stated to be the preference of engineers working on building construction projects.
Step 1 also helps with the "data checking" process in step 2. Also, if needed, new data can be collected and added to the existing CAD data. This process helps GIS managers to identify the type of utility (e.g., a sewer pipe network) and the accuracy of its location. The metadata—that is, the information that categorizes the data—need to be accurate and up to date. They indicate how, where, when, and by whom the data were collected. Metadata also compile the data assets into an inventory and provide information such as to whom they are available, their projection and coordinate system, and when they were last updated. Keeping these records will reduce duplication and will allow GIS managers to save time. For example, problems related to the misidentification of CAD data can lead to accidentally introducing errors when working in GIS. Developers and utility providers have a vested interest in assessing the accurate location of their infrastructures in spatial relation to other, public and private, infrastructures that could be co-located in underground space.

### 3.1.2. CAD Data Cleaning

The goal of step 2 is to identify and remove redundant and unnecessary information directly in CAD; this step can significantly facilitate the GIS cleaning process, part of step 5. For example, CAD maps may have data on sidewalks that may not be needed in GIS, and it may be preferable to remove the sidewalks directly in the CAD file. Annotations offer another good example as most CAD drawings contain information as text that are recognized as polylines in GIS, and it is, therefore, preferable to remove them directly from CAD files if possible. Nonetheless, instances also exist where annotations give important information about the features and thus should be included in GIS in some other form (e.g., pipe diameter that should be included in the attribute table in GIS).

In addition, because topology and geometry problems in CAD maps may be transferred in the conversion process, several problems could arise when performing spatial analyses in GIS. For example, a common topology error after converting CAD data is with polylines that do not meet perfectly at a point. It is cumbersome to carry out this process manually, especially in cases where the CAD drawings contain more information that "breaks" a polyline or polygon (see Figure 3(a)).

### 3.1.3. GIS Conversion

Step 3 tends to be a straightforward process as many software packages (including ESRI's ArcGIS) have an option to read CAD data from their GIS platform. During the conversion from CAD drawing to GIS vector data, ArcGIS, for example, divides the files into four layers: point, polyline, polygon, and annotation. Often, the points, lines, and polygons are converted into shapefiles (.shp), which is a format recognized by most GIS software packages and that store geospatial information as vectors. Annotations are not exported as shapefiles because they do not occupy space. They can, however, be manipulated as a GIS feature class in GIS.

### 3.1.4. Georeferencing

While shapefiles are created for each feature—that is, the points, polylines, and polygons—they do not have a known coordinate system. The fourth step of the process, referred to as "georeferencing", is essential because the location of each feature is assigned.

Specifically, georeferencing is a process of adding geographic information to the data so that the GIS software package can properly locate the features geographically. Many processes exist to carry out this step, and a standard process is shown in Figure 2. To carry out the georeferencing process, we need to have a shapefile (reference data) with the desired coordinate system and features like buildings that also exist in the converted CAD-to-GIS shapefiles. The converted shapefiles are then moved while keeping the reference data as a base until they match the reference shapefile. Finally, the same coordinate system of the reference data can be applied to the converted shapefiles. We should highlight here that it is crucial to ensure the geometry of the infrastructure is accurate before starting the georeferencing, hence the need to carefully carry out steps 1 and 2 first.

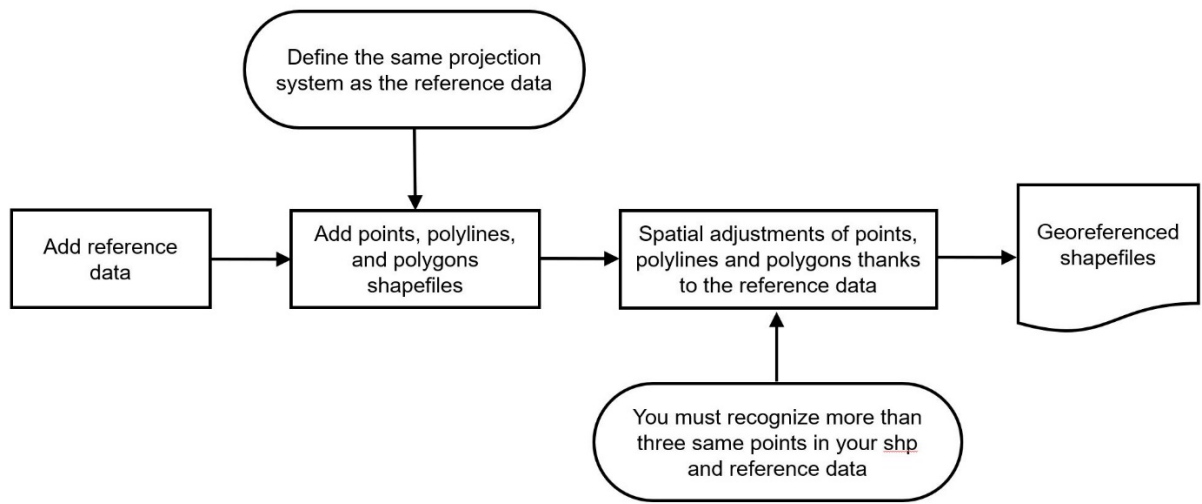

Figure 2: Georeferencing process flow chart

3.1.5. GIS Data Cleaning

Since not all the conversion issues can be addressed in CAD, some have to be addressed after the GIS conversion. This step is most often done manually, by GIS experts with knowledge of the infrastructure being converted, but some studies have attempted to develop machine learning algorithms to help with the identification of errors (Badhrudeen et al. 2020) and more work is expected in the future to help automate this process.

**3.2. Common Problems**

Table 3 lists the common problems encountered (or expected to be encountered) during the conversion process. The problems listed are not exhaustive. Some of the common problems encountered during the conversion process listed are described in more detail in this section.

| Platform | Problem description | Example |
|---|---|---|
| CAD | Insufficient (or no) metadata | No information about the infrastructure network that has abandoned pipes. |
| | Misplacement of text | Pipe diameter details placed inaccurately near another pipe. |
| | Inaccurate geometry | Wrong building shape (see below: section c) |
| | Text separating lines, thus creating gaps | (see below: section a) |
| GIS | Polygons made by continuous lines, but not closed | Buildings without a closed form. |
| | Annotations | (see below, section c) |
| | No georeferencing | (see below: section 4. d) |
| | Redundant polygons | (see below: section d) |

Table 3: List of problems in CAD and GIS

3.2.1. Texts in CAD data

Placements of texts in CAD data can create topology problems after the conversion to GIS, as illustrated in Figure 3. Texts are used in CAD to convey some information like pipe diameter, building name, street name, among others.
If the CAD data is converted into GIS, as represented in Figure 3, the space where the text '18''' is placed will create a topology problem. While these types of issues may be solved after the conversion, they are generally more easily solved directly in CAD.

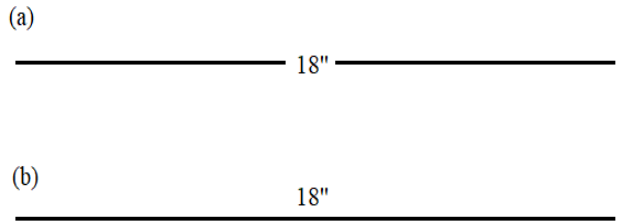

Figure 3: (a) Before correction: text in the middle of the line
(b) After correction: text above the line

### 3.2.2. Annotations conversion

Relevant information may be present in the form of text in CAD data that should also be included in GIS. This can be done by converting the text into annotations in GIS and exporting them as a feature class that becomes part of the geodatabase. After converting annotations into a feature class in GIS, a point feature is created and starts to serve as a proxy that specifies the location for the text, which can then be exported as a shapefile. In other words, annotations in CAD are converted into points in GIS and assigned to a specific layer and stored as an attribute. More specifically, they can be preserved by transferring them into the attribute table to the nearest point, polyline, or polygon, as shown in Figure 4.

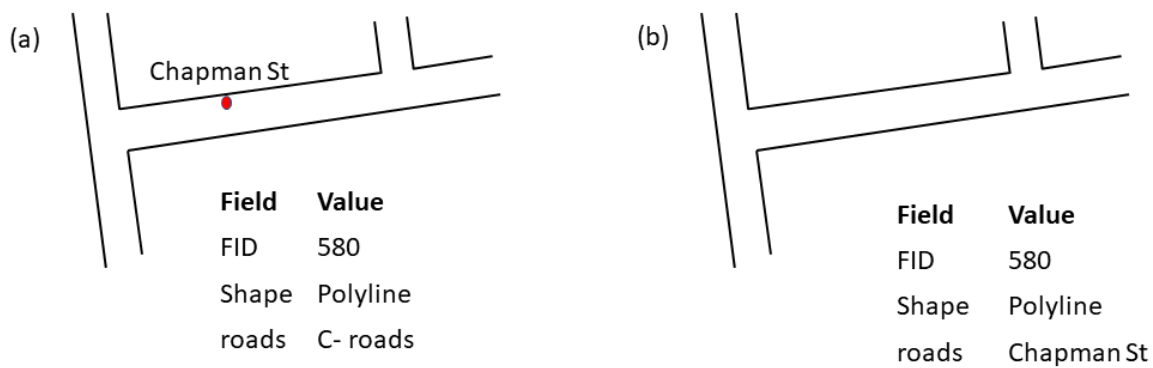

Figure 4: (a) CAD annotation (b) line (with the attribute) in GIS

### 3.2.3. Inaccurate geometry

Problems may arise during the georeferencing step when the geometry and measurements of the buildings are inaccurate. This can create problems when trying to overlay the CAD data correctly over the reference GIS data. For example, in Figure 5, if the building in the CAD data (green line) is not accurate, then we cannot have a perfect overlay on the reference data (solid green block), and it, therefore, becomes difficult to properly and accurately georeference the data.

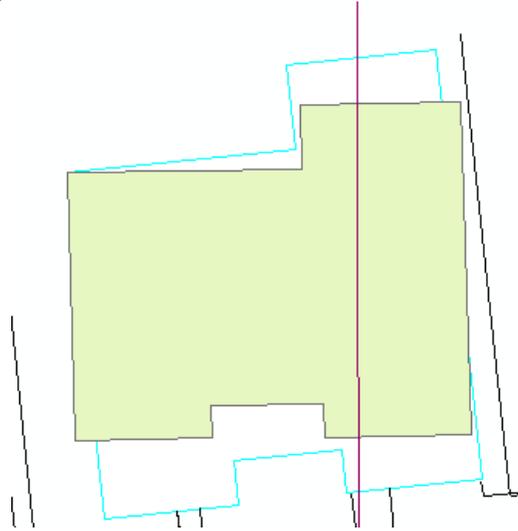

Figure 5: Wrong building shape complicating the georeferencing of the data.

3.2.4. Redundant polygons

Blocks and lines sometimes represent single entities in CAD and therefore need to be converted to points in GIS. For example, in Figure 6, manholes are represented as circles in CAD, whereas they should be represented by points in GIS.

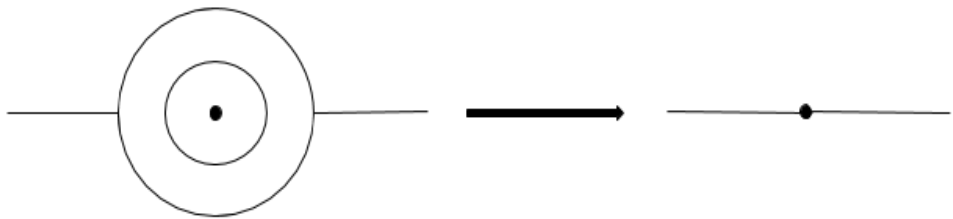

Figure 6: Manholes to a point feature in GIS

## 4. Case Study

The five-step process explained in this work is applied to an underground wastewater system provided by the University of Illinois at Chicago (UIC) Office of Capital Planning & Project Management (OCPPM). This system covers the UIC west campus. The main goal is to convert the CAD drawing data of the underground pipe network into a GIS format that contains different shapefiles for elements such as manholes, catch basins, and conduits. Figure 7 shows the CAD data used for this case study. The conduits to be converted to GIS are shown in pink and green.

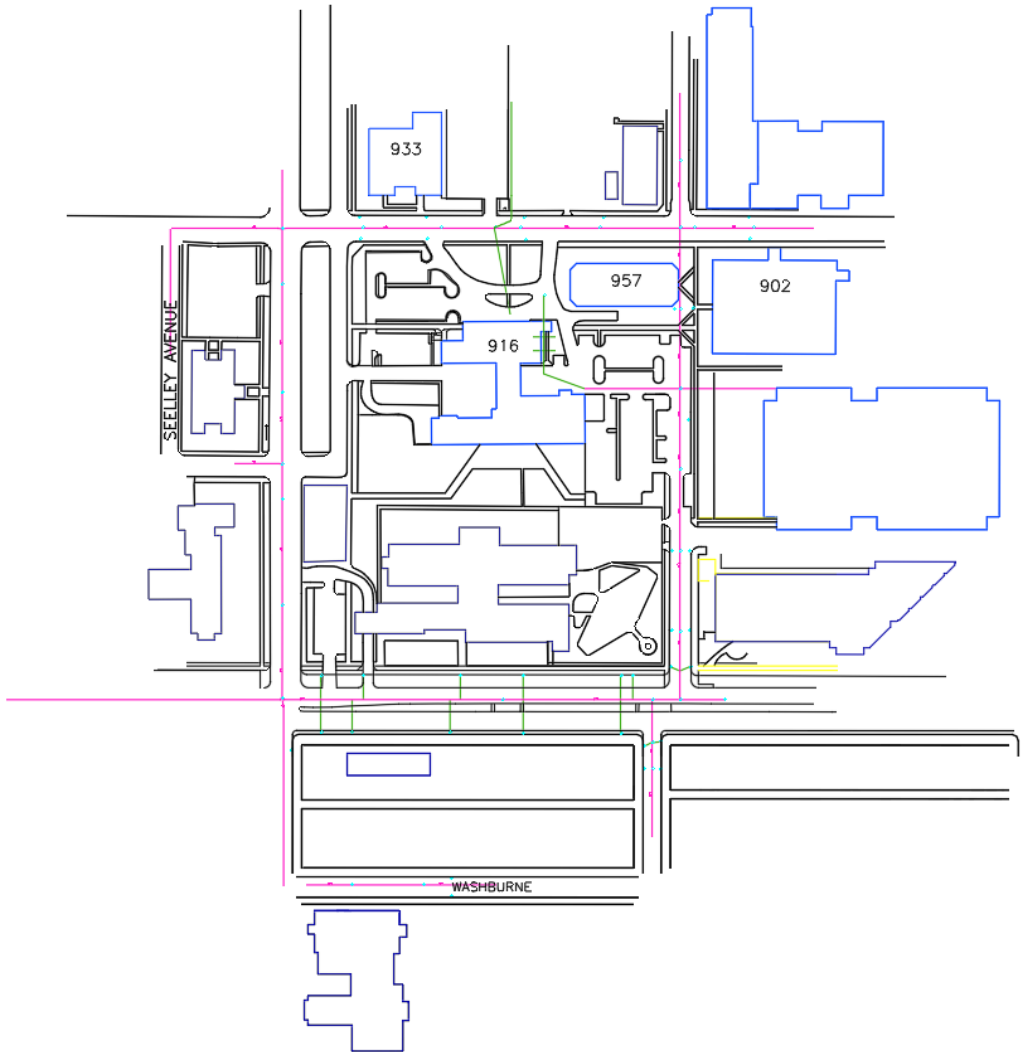

Figure 7: Wastewater system map of the UIC West campus

The underground wastewater system consists of a main sewer conduit located in the road that is connected to smaller conduits that collect wastewater from buildings and from stormwater catch basins; Chicago has a combined sanitary and stormwater sewer system. In addition to the stormwater catch basins, manholes are present to give access to the main sewer conduit in the road. The important information to collect for this type of system are: location of manholes, catch basins, and conduits that connect the catch basins. For this project, we convert the locations of the manholes and the catch basins, as well as the location of all wastewater conduits, and connect them in GIS.

### 4.1. Step 1: Identification of Needs

For this case study, the first step was omitted because the UIC OCPPM was able to define for themselves the data they required and kept that data up to date. Nevertheless, for illustrative purposes, if we were to proceed with the first step, we would first collect information such as distance between manholes and a benchmark, and randomly check the distances between two catch basins. This would provide us with some relevant information to assess the CAD data accuracy.

### 4.2. Step 2: CAD Data Cleaning

The CAD drawing contained some irrelevant information for this case study, such as the presence of sidewalks, as would have been identified in step 1. Depending on the needs of the particular municipality, irrelevant information can be ignored for the conversion to GIS. Most of this data can be deleted from the CAD files directly. In contrast, other information needed to be retained, such as data on roads or buildings that will be used for georeferencing.

### 4.3. Step 3: GIS Conversion

Converting the CAD data into shapefiles is a straightforward process. ArcGIS projects the CAD data automatically, even without any coordinate system. Figure 8 shows the projected CAD data in the ArcGIS platform.

In Figure 8, a list of feature classes is shown in the table of contents on the left-hand side, including point, polyline, polygon, multipatch, et cetera. Since the necessary information that needs to be converted into shapefiles are conduits (i.e., polylines), manholes (points), and storm catch basins (polygons), they can be selected and exported as a shapefile. Nonetheless, it should first be georeferenced, which is the goal of the next step.

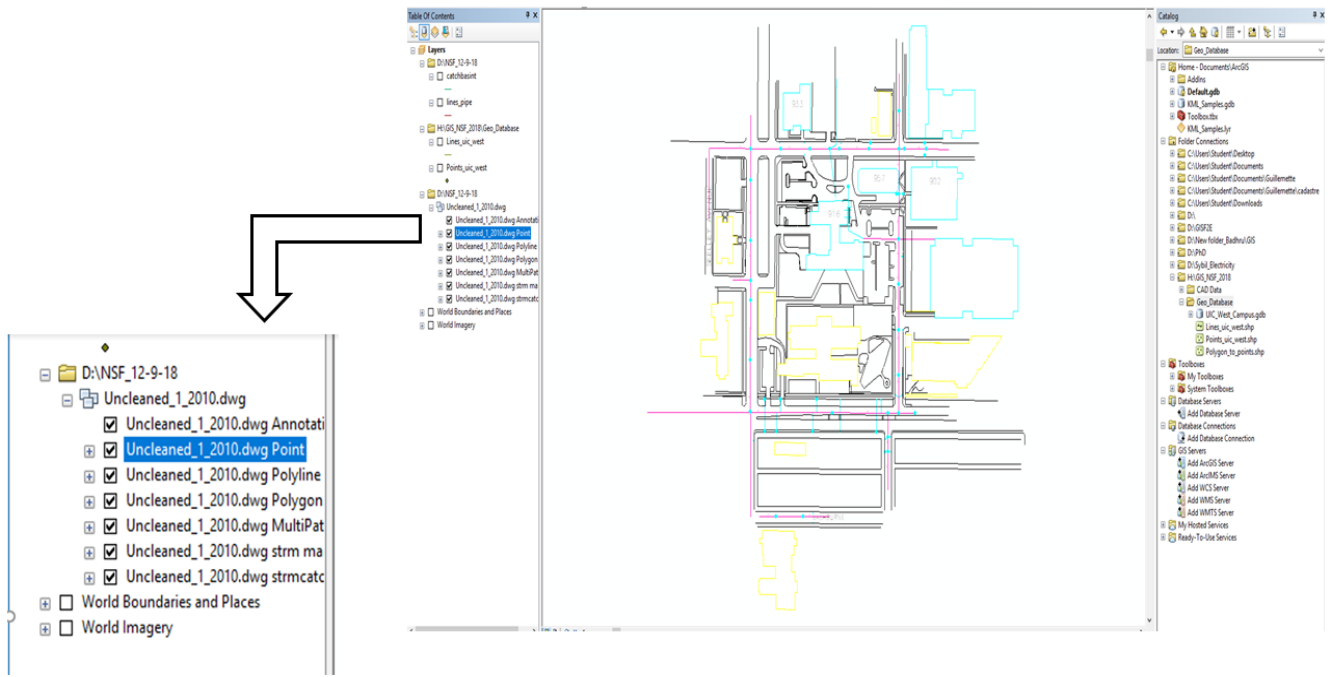

Figure 8: CAD projected in ArcGIS

## 4.4. Step 4: Georeferencing

The toolbar in ArcGIS has a tool named "Georeferencing" that can be used to assign the geographic position information to the CAD data. Figure 9 shows the original position of the data on a world map, essentially in the Atlantic Ocean, South of West Africa, at coordinate zero for both the longitude and the latitude. During the georeferencing step, the CAD data can be manipulated, for example, by shifting, rotating, and scaling it to make it fit perfectly on the reference map. Here, we use the raster image of the world map, but any other properly georeferenced GIS data can be used. Nevertheless, despite trying a significant number of configurations, some spatial distortions in the converted data persist, and all CAD data, therefore, cannot fit over the world map perfectly. Figure 9 (b) shows the georeferenced CAD data, and as it can be seen, the overlay is not perfect as the shapes of the buildings are not accurate.

Once the georeferencing is completed as properly and accurately as possible, the data can be exported as shapefiles, and the shapes are converted into points, polylines, and polygons.

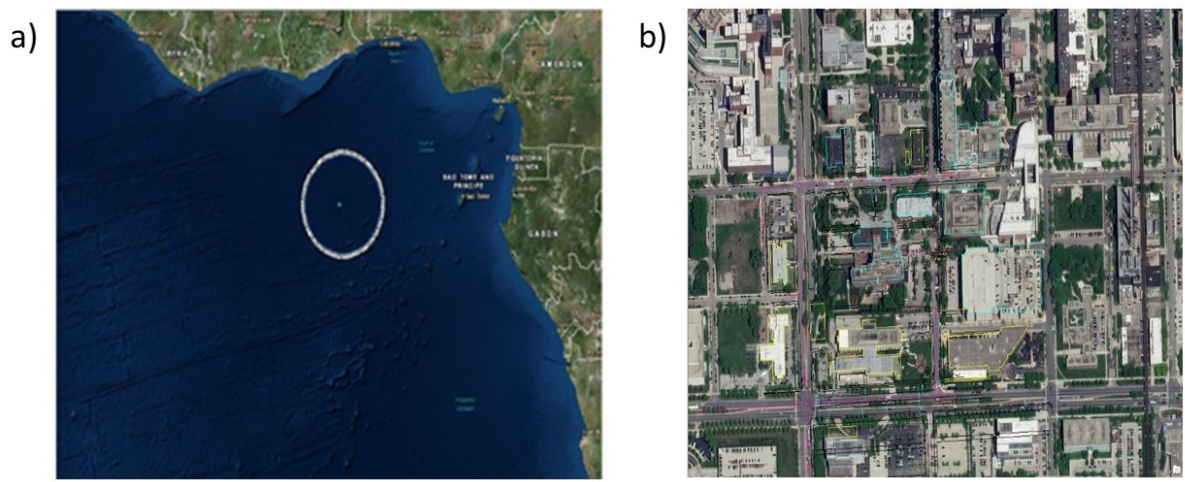

Figure 9: Before (a) and after (b) georeferencing of the CAD data

## 4.5. Step 5: GIS Data Cleaning

As mentioned above, GIS cleaning is often carried out manually. In this case study, catch basins and manholes presented a problem because they were converted into polygons. We would prefer to have them converted into points. Various ad hoc processes, some of which can be automated, then need to be implemented to clean the GIS data as properly as possible.

# 5. Conclusions

Many municipalities and organizations around the world are converting and aspire to convert CAD drawings to GIS to enable the types of assessments and analyses that can be performed in GIS. Nonetheless, the existing conversion process is time-consuming and requires users to be knowledgeable in both CAD and GIS. Many municipalities have limited budgets and capacity of staff trained in GIS, which makes the time spent on the conversion process even more valuable. Exploratory interviews with municipal GIS analysts and managers across the U.S. found that this type of data conversion was typically performed by those who were proficient in either CAD or GIS and that they would benefit from a standardized conversion protocol. A step by step process would allow anyone with a basic knowledge of CAD and GIS to convert data in a timely manner without compromising accuracy.

In response to this need, this article presented a generalized process for the conversion of data from CAD drawings to GIS shapefiles. The generalized process presented here was validated by its application to both the case of UIC data as well as its ability to resolve the conversion problems raised by the interviewees. To summarize, the process has five steps: identification of needs, CAD data cleaning, GIS conversion, georeferencing, and GIS data cleaning. These steps were demonstrated with the conversion of actual CAD data into GIS shapefiles.

Future research could explore generalizing this process to other network systems and identifying the steps that can be automated to reduce further the time required by the conversion process. This would further enable those working with GIS in municipalities and other organizations to fully and effectively use their GIS-based infrastructure data.

## 6. Funding

This research is partly supported by the National Science Foundation (NSF) CAREER award [#155173] and by the NSF Cyber-Physical Systems (CPS) award [#1646395].